\documentstyle[epsfig,aps,preprint]{revtex}
\begin{document}
\draft
\preprint{\vbox{To appear in Physical Review {\bf C}\hfill
        \hfill USC(NT)-98-04}}
\tolerance = 10000
\hfuzz=5pt
\title{On the chiral low-density theorem}
\author{V. Dmitra\v sinovi\' c}
\address{Department of Physics and Astronomy,\\
University of South Carolina, Columbia, SC 29208, USA} 
\date{\today}
\maketitle
\begin{abstract}
We show how the linear ``low-density theorem" of Drukarev and Levin can 
be extended to arbitrary positive integer power of the baryon density $\rho$.
The $n^{\rm th}$ coefficient in the McLaurin expansion 
of the fermion condensate's $\rho$  dependence is 
the connected $n$-nucleon sigma term matrix element. We calculate the 
$O(\rho^2)$ coefficient in lowest-order perturbative approximation to the 
linear sigma model and then show how this and other terms can be iterated to
arbitrarily high order. Convergence radius of the result is discussed.
\end{abstract}
\widetext

\section{Introduction}

Earlier this decade a ``low-density theorem" (LDT) for the linear term 
in the baryon density dependence of the quark condensate was formulated 
\cite{dl91,shak92,cfg92,weise92}. This first term 
in the density power series expansion of the quark condensate 
is proportional to
the nucleon/constituent quark $\Sigma$ term, which leads one to believe 
that, perhaps, it is only the chiral symmetry breaking of the 
(strong) hadronic interactions by the current quark masses that 
controls the baryon density dependence of the quark condensate 
and thus makes it essentially unique.
Soon thereafter it became clear, however,
that the extrapolation of this linear formula from the low-density region 
upwards, to the neutron star densities, would be highly contentious. 

In this paper: (1) We present an
extension of the Drukarev-Levin low-density theorem to terms of
arbitrarily high order in the McLaurin expansion of the fermion condensate's
dependence on the baryon density $\rho$. The n-th McLaurin coefficient is just 
the connected n-nucleon $\Sigma$ term (elastic matrix element), 
which is a function of 
the explicit chiral symmetry breaking ($\chi$SB) terms in 
the Hamiltonian of the theory.
(2) We show how the ``theorem" works in an explicit example:  We calculate the 
$O(\rho^2)$ coefficient in the lowest perturbative approximation to the 
linear sigma model and then iterate these diagrams to infinite order.
The result can be resummed and we obtain a closed-form solution - we 
use this result to establish the range of validity of such a diagrammatic 
calculational scheme.

\section{Baryon density dependence of the $\Sigma$ term}

\subsection{Proposition}

We propose an extension of the
low-density theorem of Drukarev and Levin \cite{dl91}
to arbitrary (positive integer) powers of the 
baryon number/quark density $\rho$: 

The $n^{th}$ coefficients in the McLaurin 
expansion of $\langle \Sigma \rangle_{\rho}$ is the
n-fermion/nucleon/quark 
matrix element of the (pion) $\Sigma$ double commutator 
\begin{eqnarray} 
\Sigma 
&=&  
{1 \over 3} \sum_{a = 1}^{3} \big[Q^{a}_{5},
[Q^{a}_{5}, {\cal H}_{\rm \chi SB}(0)]\big] 
~,\
\label{e:sig}
\end{eqnarray}
is
\begin{eqnarray} 
\langle \rho \left| \Sigma \right| \rho \rangle 
&=&  
\sum_{j = 0}^{\infty}
{\rho^{j} \over{j!}} 
\langle j N \left| \Sigma \right| j N \rangle_{\rm connected}
~,\
\label{e:theorem}
\end{eqnarray}
where the sum over $j$ involves {\it connected matrix elements}.

As preconditions we assume that: 
(i) the color symmetry of the theory is neither
spontaneously nor explicitly broken;
(ii) the theory can be described by a Lorentz invariant, local,
approximately chiral-invariant Lagrangian density with a
partially conserved axial N\" other current;
(iii) the axial anomaly does not invalidate the 
relevant axial Ward identities.

The proof is based on the current-algebraic relation/identity: 
\begin{eqnarray} 
\langle \Sigma \rangle_{\alpha} &=&  
\langle \alpha \left| \Sigma \right| \alpha \rangle
= {1 \over 3} \sum_{a = 1}^{3} \langle \alpha \left| 
\big[Q^{a}_{5}, [Q^{a}_{5}, {\cal H}_{\rm \chi SB}(0)]\big] 
\right| \alpha \rangle
\nonumber \\ 
&=&  
- {1 \over 3} f_{\pi}^{2} \lim_{k \to 0} \sum_{a = 1}^{3} 
\langle \pi^{a}(k) \alpha \left| S \right| \pi^{a}(k) \alpha \rangle
~,\
\label{e:sterm}
\end{eqnarray} 
between the exact elastic soft-pion
$\pi \alpha$ scattering (S) matrix and the corresponding 
Heisenberg representation (pion) sigma term matrix element.
Here $a$ is the flavour index of the external pions, which is
averaged over the three varieties of pions, i.e., $a = 1, 2, 3$.
This result can be adapted to the kaon sigma term by specifying 
$a$ = 4, 5, 6, or 7, or by averaging over some subset thereof.
Formula (\ref{e:sterm}) is 
derived from a chiral Ward identity and the LSZ reduction formula 
for the case of a nucleon ($\alpha = N$) on p. 131 - 7 of Ref. \cite{sak69}.
[This is equivalent to applying Sakurai's ``master formula'', p. 111 - 2 
in Ref. \cite{sak69}, 
twice, once to the initial and once to the final state.]
Note that the sigma term is only sensitive to 
the chiral symmetry breaking ($\chi$SB) terms in the 
Hamiltonian density ${\cal H} = {\cal H}_{\rm \chi} +
{\cal H}_{\rm \chi SB}$
\begin{eqnarray} 
\big[Q^{a}_{5}, [Q^{a}_{5}, {\cal H}]\big] 
&=&  
\big[Q^{a}_{5}, [Q^{a}_{5}, {\cal H}_{\rm \chi SB}]\big] 
,\
\label{e:ster1}
\end{eqnarray} 
the chiral charges being constants of the motion in the chiral
limit
\begin{eqnarray} 
\left[Q^{a}_{5}, {\cal H}_{\rm \chi} \right] 
&=&  0 .\
\label{e:ster2}
\end{eqnarray} 

Now apply this equation to the nuclear matter state ket
$| \alpha \rangle = | \rho \rangle = | n N \rangle$.
We shall work in a large but finite box of volume $\Omega$
in order to avoid momentum-conserving Dirac delta functions, 
which are replaced by Kronecker ones. 
This also means that we can
have a finite baryon density $\rho = {n \over \Omega}$ without 
taking the number of baryons to infinity, $n \to \infty$, which
we leave for the very last step. Therefore
\begin{eqnarray} 
\langle \Sigma \rangle_{\rho}
&=&  
\langle \rho \left| \Sigma \right| \rho \rangle
\nonumber \\  
&=& 
\lim_{n = \rho \Omega \to \infty}
\langle n N \left| \Sigma \right| n N \rangle
\nonumber \\ 
&=& 
\lim_{n = \rho \Omega \to \infty}
{1 \over 3} \sum_{a = 1}^{3} \langle n N \left| \big[Q^{a}_{5},
[Q^{a}_{5}, {\cal H}_{\rm \chi SB}(0)]\big] \right| n N \rangle
\nonumber \\ 
&=&  
- {1 \over 3} f_{\pi}^{2} 
\lim_{n = \rho \Omega \to \infty}
\lim_{k \to 0} \sum_{a = 1}^{3} 
\langle \pi^{a}(k) n N \left| S \right| \pi^{a}(k) n N \rangle
~,\
\label{e:def}
\end{eqnarray}
where we used the fact that the finite density (zero temperature)
ground state wave function is just the wave function of $n$ 
fermions/nucleons at rest enclosed in a box of volume $\Omega$.
Formula (\ref{e:def}) relates the finite density $\Sigma$ term 
matrix element to the {\it exact} 
(forward) scattering amplitude of a {\it soft}, i.e.,
vanishing four-momentum ($k = 0$) pion from $n$ nucleons at rest.

Here we have merely used the definition of the $\Sigma$-term to rewrite
the object of interest in terms of an S-matrix element. To be sure, no
new information was gained in this step and at no loss of generality.
The result is that we may now use the general results of S-matrix theory. 
That, in turn, allows a Feynman-diagrammatic organization of our problem.  
The most important property of the S-matrix in this regard is its
decomposability into disconnected and connected parts. The latter part
falls further 
into reducible and one-, two-, three-particle, etc. irreducible classes. 
Specifically, in our case of pions and nucleons we may separate the exact 
(Heisenberg representation) S-matrix into the following distinct categories: 
(i) the completely disconnected graph (one pion and $n$ baryons 
all propagating without interaction); 
(ii) $n$ simply connected $\pi N$ scattering amplitudes multiplied by 
$(n - 1)$ disconnected baryon lines;
(iii) $\left(\begin{array}{c} n \\ 2 \end{array} \right) 
= {n (n - 1) \over 2}$ simply connected 
$\pi N N$ scattering amplitudes multiplied by $(n - 2)$ 
disconnected baryon line, {etc.} \cite{hub57}. Each of the disconnected
subdiagrams has its own momentum-conserving delta function, 
which translates into one volume ($\Omega$) factor for 
each disconnected baryon line
\footnote{This, of course, means that the
dimension of the various $\Sigma$ term matrix elements
varies with the number of nucleons 
(baryons/quarks more generally):
the vacuum $\Sigma$ term has dimension $M^4$, the
(single-)nucleon $\Sigma$ term has dimension $M^1$, etc., i.e., 
$M^{4 - 3 n}$ for the $n$-nucleon $\Sigma$ term} \cite{hub57}. 
Therefore 
\begin{eqnarray} 
\langle n N \left| \Sigma \right| n N \rangle
&=&  
- {1 \over 3} \sum_{a = 1}^{3} f_{\pi}^{2} \lim_{k \to 0} \Bigg[
\langle \pi^{a}(k) \left|~ S~  \right|  \pi^{a}(k) \rangle
\langle n N | n N \rangle
\nonumber \\ 
&+&  
~~~n~~ 
\langle \pi^{a}(k) N \left|~ S~ \right| \pi^{a}(k) N
\rangle _{\rm connected}~~~
\Omega^{-1} \langle n N | n N \rangle
\nonumber \\ 
&+&  
\left(
\begin{array}{c} 
n \\ 2 
\end{array}
\right) 
\langle \pi^{a}(k) (2 N) \left|~ S~  \right| \pi^{a}(k) (2 N) 
\rangle_{\rm connected}
\Omega^{-2} \langle n N | n N \rangle + ... \Bigg]
\nonumber \\  
&=&  
- {1 \over 3} \sum_{a = 1}^{3} f_{\pi}^{2} \lim_{k \to 0} 
\sum_{j=1}^{n}
\left(
\begin{array}{c} 
n \\ j 
\end{array}
\right)  \Omega^{-j} 
\langle \pi^{a}(k) (j N) \left|~ S~  \right| \pi^{a}(k) (j N)
\rangle_{\rm connected} 
\langle n N | n N \rangle
~.\
\label{e:sig1}
\end{eqnarray}
Now turn this back into a statement about the $\Sigma$ terms
\begin{eqnarray} 
{\langle n N \left| \Sigma \right| n N \rangle 
\over{\langle n N | n N \rangle}}
&=&  
\sum_{j = 0}^{n}
\left(
\begin{array}{c} 
n \\ j 
\end{array}
\right) 
\left({1 \over \Omega}\right)^{j} 
\langle j N \left| \Sigma \right| j N \rangle_{\rm connected}
~. \
\label{e:res1}
\end{eqnarray}
We are now ready to take the thermodynamic limit 
$\Omega \to \infty$ at (finite) constant density $\rho$.
Since we assumed a large box to begin with, the baryon number 
$n = \rho \Omega$ contained therein may be taken
to be far larger than any given order in the power expansion, $n \gg j$,
and in the thermodynamic limit $n \to \infty$ one may write 
\begin{equation} 
\left(
\begin{array}{c} 
n \\ j 
\end{array}
\right)  = {n! \over{j! (n - j)!}}
\simeq {n^{j} \over{j!}},
\end{equation}
which leads to the announced result
\begin{eqnarray} 
\langle \rho \left| \Sigma \right| \rho \rangle 
&=&  
\lim_{n = \rho \Omega \to \infty} \sum_{j = 0}^{n}
{1 \over{j!}} \left({n \over \Omega}\right)^{j} 
\langle j N \left| \Sigma \right| j N \rangle_{\rm connected}
\nonumber \\ 
&=&  
\sum_{j = 0}^{\infty}
{\rho^{j} \over{j!}}  
\langle j N \left| \Sigma \right| j N \rangle_{\rm connected}
~, ~~~~~~~{\rm q.e.d.}\
\label{e:res2}
\end{eqnarray}
when the nuclear ground state is 
normalized $\langle \rho | \rho \rangle = 1$.
It is manifest that this result does not hold in {\it finite} 
systems at orders $j$ comparable to the total baryon number $n$.
 
A reminder seems in place that this series need {\it not} 
uniquely determine $\langle \Sigma \rangle_{\rho}$ for arbitrary
values of $\rho$ because its radius of convergence 
may be small, or even zero. One possible reason for such a behaviour  
of a power series is that it really is a Laurent one, 
i.e., the function may have poles and/or branch cuts in the complex baryon
density ($\rho$) plane. Explicit calculations will show this to be the case.

\subsection{Consequences}

Note that the ``mechanical" sigma operator 
\begin{equation} 
\Sigma_{\rm mech} = 
{1 \over 3} \sum_{a=1}^{3} \big[Q^{a}_{5},
[Q^{a}_{5}, {\cal H}_{\rm mech}(0)]\big] 
= {\cal H}_{\rm mech}(0), 
\end{equation}
equals the ``mechanical'' $\chi$SB Hamiltonian 
$${\cal H}_{\rm mech} = {\bar \Psi} m_{q}^{0}\Psi =
m_{u}^{0} {\bar u}u + m_{d}^{0} {\bar d}d, $$ 
due to the current quark masses, or the bare nucleon mass term 
$${\cal H}_{\rm mech} = {\bar N} M_{N}^{0}N=
M_{p}^{0} {\bar p}p + M_{n}^{0} {\bar n}n . $$ 
Our result Eq. (\ref{e:res2}) holds either for quarks, or for baryons as the 
fermions.
Since the chiral symmetry breaking is defined in terms of quark degrees 
of freedom, but we do not know the solution to the quark dynamics in baryons,
we must first establish, or rather postulate a relation between 
the quarks' and the observable hadrons' properties. 
One assumption common in the literature is that 
$\Sigma_{N} = 3 \sigma_{Q}$.
\footnote{ 
It is clear that this amounts to the ``impulse approximation",
but it was recently shown that 
the two-quark operator corrections {\it must} exist \cite{sato98}
if chiral symmetry is to be preserved.}
Assuming, by the same token, effective proportionality of the nucleon and 
quark condensates, our Eq. (\ref{e:res2}) leads to a statement about 
the quark condensate at finite baryon density
\begin{eqnarray} 
{\langle {\bar N} N \rangle_{\rho}
\over{\langle {\bar N} N \rangle_{0}}} =
{\langle {\bar q} q \rangle_{\rho}
\over{\langle {\bar q} q \rangle_{0}}}
&=&  
1 - {1\over{(f_{\pi} m_{\pi})^{2}}}
\sum_{j = 1}^{\infty} {\rho^{j} \over{j!}} 
\langle j N \left| \Sigma \right| j N \rangle_{\rm connected} 
~. \
\label{e:res3}
\end{eqnarray}
The $j = 1$ term together with the unity on the right-hand side form the
Drukarev-Levin LDT. The rest of this formula appears to be new \footnote{
The idea that the elastic $\pi-n N$ scattering amplitude is related to the
$\rho$-dependence of $\langle {\bar q} q \rangle_{\rho}$ is implicit in 
Refs. \cite{eric93,birse93,lyon95}.}.

This form of the LDT has been used to argue about 
the behavior of the p.s. meson masses at non-zero baryon density. We
remind the reader that the quark condensate is not the only
density-dependent variable influencing the p.s. meson mass - one 
must take into account the (model dependent) p.s. meson decay constant's 
density dependence as well, which is not commonly done.
Lest this result leads one to believe that the problem of the
density dependence of p.s. meson masses is solved, we remind the reader 
that neither the radius of convergence of the series is known, nor 
do we have, as yet, an efficient 
algorithm for calculating the higher-order coefficients in the expansion.

We next present an
example of a perturbative calculation in the (chiral) linear sigma model
and its iteration to arbitrarily high order. 

\section{An example}

Before we proceed we remind the reader of the assumptions of the LDT: 
the matrix elements must be either the 
exact Heisenberg representation ones, or approximate matrix elements 
that satisfy the underlying chiral Ward identity (master formula). 
The former, is beyond our powers, 
the latter are available, mostly in the form of 
perturbative solutions, though solutions to several models have been
found that sum certain infinite classes of Feynman diagrams 
(we shall not consider those approximations here).
We shall calculate the terms of $O(\rho^2)$ in the chiral linear sigma model, the 
$O(\rho)$ calculations having been done before \cite{shak92,cfg92,weise92}.
This calculation will be with (free) nucleon degrees of freedoms, new methods for 
calculation of sigma terms in interacting nucleon systems having been developed 
only recently \cite{sato98}.

\subsection{One- and two-nucleon sigma terms in the linear sigma model}

The one-nucleon connected sigma term can be extracted from the elastic $\pi$N 
scattering amplitude, which in the linear $\sigma$ model is given by the sum of 
the three diagrams in Fig. 1: two nucleon-pole diagrams, (b) and (c), and 
one $\sigma$-exchange graph (a). These have been calculated many times 
\cite{camb79} and will not be repeated here. 

The pion-two-nucleon connected matrix element is substantially more complex:
there are more than 50 connected diagrams in the lowest-order
perturbative (Born) approximation. All of them are, by definition, reducible
diagrams. We separate these graphs into three distinct subsets:
1) pion ``rescattering''; 2) pion scattering on a sigma-meson-in-flight;
3) initial-, final-, and intermediate-state interaction diagrams. 

1) Eighteen ($2 \times 3 \times 3$) of these graphs form the simplest connected, 
reducible ``pion-rescattering'' amplitude that is built up from two 
pion-one-nucleon effective vertices, Fig. 1. Their contribution equals
\begin{eqnarray}
\langle \Sigma \rangle_{2N}^{\pi} 
&=& 
- 2 \left({\langle \Sigma \rangle_{N} \over{f_{\pi} m_{\pi}}}\right)^{2}~. \
\label{e:sigma2n1}
\end{eqnarray}
The currently accepted value of the nucleon sigma term
$\langle \Sigma \rangle_{N}$ lies between 45 and 65 MeV. 

2) Only the sigma terms of scalar states, discrete or continuum, exchanged 
between the two nucleons contribute to the two-nucleon sigma term. 
[This follows from the fact that the sigma term is a Lorentz scalar.]
In the first Born approximation to the linear sigma model the only such object 
is the sigma meson, so we end up having to calculate its sigma term and its 
contribution to the the two-nucleon sigma term.
Four Feynman diagrams, see Fig. 2, contribute to the elastic pion-sigma meson 
scattering amplitude and hence to also to its sigma term. The result is 
\begin{eqnarray}
\langle \sigma | \Sigma | \sigma \rangle 
&=& 
3 m_{\pi}^{2}~. \
\label{e:sigma12}
\end{eqnarray}
Inserting this result into the two-nucleon sigma term we find
\begin{eqnarray}
\langle \Sigma \rangle_{2N}^{\sigma} 
&=& 
3 g_{0}^{2} \left({m_{\pi}^{2} \over{m_{\sigma}^4}}\right)~, \
\label{e:sigma2n}
\end{eqnarray}
where $g_{0} = \left(g_{A} M/ f_{\pi}\right) = 12.6$ is the 
$\sigma$-nucleon coupling constant.
The currently accepted value of the $\sigma$ mass lies between
400 MeV and 1200 MeV \cite{pdg98}.

3) These are {\it not all} of the reducible graphs at this order, however: 
all of the remaining graphs can be described as either initial-, intermediate-,
or final-state interactions, see Fig. 3. They are formed either by having a 
complete
pion-nucleon scattering amplitude attached to one external nucleon line,
or by having a pion-nucleon vertex attached to two external nucleons
in the sum of one meson exchanges between the two nucleons (the NN potential).
This one-meson exchange potential ought to be summed into an
infinite ladder by iteration of the Bethe-Salpeter equation and thus made to
form a proper interacting NN (scattering or bound) state. In the end,
this summation is equivalent to the introduction of NN correlations
into the nuclear matter.
These diagrams diverge {\it in vacuo}, although they are Born diagrams, because the
intermediate nucleons are on their mass shells, due to the external soft-pion 
condition.

Points 1) and 2) put together lead to
\begin{eqnarray}
\langle \Sigma \rangle_{2N}  
&=& 
\langle \Sigma \rangle_{2N}^{\pi} +
\langle \Sigma \rangle_{2N}^{\sigma}
\nonumber \\
&=& 
- 2 \left({- \langle \Sigma \rangle_{N} \over{f_{\pi} m_{\pi}}} \right)^{2}
+ 
3 \left({g_{0} m_{\pi} \over{m_{\sigma}^{2}}}\right)^{2}~, \
\label{e:sigma2nt}
\end{eqnarray}
for the total. 
With the currently accepted value of the $\sigma$ mass, 
the second term exceeds the first one.

One feature of this result stands out:
the appearance of the second power of the single-nucleon sigma term coming from
the repetition of the one-pion reducible diagrams involving the ``primitive'' 
(elementary) single-nucleon sigma term contributions. 
It has been pointed out by M. Ericson \cite{eric93} that this sort of behaviour 
generalizes to higher orders of perturbation theory, Fig. 4, and that the graphs 
can be resummed into a geometric progression. We shall address that question next. 

\subsection{Iteration of primitive contributions in the N-nucleon sigma term}

In the foregone analysis one will have noticed that certain lower-order 
graphs are repeated in the higher ones with simple regularity. 
The multiplicities of these graphs can be figured out and the infinite
(sub-)series resummed. We shall show that several classes of such diagrams.
sum into geometric series. That result is particularly useful because its
properties, such as the radius of convergence and the analytic structure
are well known. Yet, as we shall show, uncritical application may sometimes
lead to erroneous conclusions.

\paragraph*{Single-nucleon pion rescattering}
The connected but reducible diagrams, Fig. 4,
consisting of individual $\pi N$ elastic scattering amplitudes
connected by a single pion line carrying zero four-momentum 
can be written as
\begin{eqnarray}  
\langle n N \left| \Sigma \right| n N \rangle_{\rm rescatt.}
&=&  
n! \langle \Sigma \rangle_{N}^{n} 
\left(- m_{\pi}^{2} f_{\pi}^{2} \right)^{(1-n)}~,\
\label{e:sig3}
\end{eqnarray}
where the factor ${1 \over{- m_{\pi}^{2}}}$ comes from the
propagation of a soft $\pi$, the 
$\langle \Sigma \rangle_{N} f_{\pi}^{-2}$ comes from a single
irreducible elastic $\pi N$ scattering amplitude (vertex) for {\it soft} 
pions, and $n!$ is the number of identical graphs.
Hence the infinite series of such connected one-pion-reducible graphs is
readily summed up as
\begin{eqnarray} 
{\langle {\bar N} N \rangle_{\rho}
\over{\langle {\bar N} N \rangle_{0}}} \Bigg|_{\rm rescatt.}
=
{\langle {\bar q} q \rangle_{\rho}
\over{\langle {\bar q} q \rangle_{0}}} \Bigg|_{\rm rescatt.}
&=&  
1 + \sum_{n = 1}^{\infty}   
\left({- \rho 
\langle \Sigma \rangle_{N} \over{(f_{\pi} m_{\pi})^{2}}} \right)^{n}
\nonumber \\
&=&  
\left[1 + 
{\rho \langle \Sigma \rangle_{N} \over{(f_{\pi} m_{\pi})^{2}}} \right]^{-1}~. \
\label{e:ldt4}
\end{eqnarray}
Since this is a geometric progression, the series has a finite radius of
convergence (in $\rho$) given by 
$$\rho \leq |\rho_{c_{1}}| = 
{(f_{\pi} m_{\pi})^{2} \over{\langle \Sigma \rangle_{N}}},$$
which equals roughly twice (1.8), or three times the normal nuclear density 
for $\langle \Sigma \rangle_{N}$ = 65, 45 MeV, respectively. 

It seems obvious that the complete series including all diagrams, at best 
can have the convergence properties of the worst-behaved sub-series. In
other words, our resummation of the one-nucleon pion rescattering diagrams 
seems to tell us the maximum reliable density calculable with such diagrammatic 
methods. 
An important practical consequence of the above reasoning and of the empirical 
value of the kaon-nucleon sigma term $\langle \Sigma \rangle_{K N}$ = 200 MeV 
is that {\it the radius of convergence of the series relevant to
the $K$-condensation is at best a fraction of one normal nuclear density}. 
Surprisingly, this conclusion turns out to be only qualitatively correct, 
due to cancellations from other two-nucleon sigma term contributions.
We shall explicitly show that in the
linear sigma model the inclusion of another two-nucleon
sigma term into the infinite series leads to an increase in the number of 
poles, change of their 
positions and to an increase in the radius of convergence of the series.

\paragraph*{Two-nucleon pion rescattering}
The connected two-nucleon ``pion-rescattering'' graphs fall
into one of the two categories: (i) even-n, and (ii) odd-n, 
and will be dealt with accordingly.

(i) In the former case, one part of the amplitude is in the form of
a product of $n/2$ pion-two-nucleon graphs connected by 
$n/2$ pion propagators. Since one can select the first pair in 
$\left(\begin{array}{c} n \\ 2 \end{array} \right)$ many 
ways, the second pair in 
$\left(\begin{array}{c} n - 2 \\ 2 \end{array} \right)$ many 
ways, etc., one finds that
\begin{eqnarray}  
\langle n N \left| \Sigma \right| n N \rangle |^{n-{\rm even}}
&=&  
- \left(\begin{array}{c} n \\ 2 \end{array} \right) 
\left(\begin{array}{c} n - 2 \\ 2 \end{array} \right) 
\cdots
\left(\begin{array}{c} 2 \\ 2 \end{array} \right) 
\left({- \langle \Sigma \rangle_{2 N} \over{m_{\pi}^{2}f_{\pi}^{2}}}
\right)^{n \over 2} (f_{\pi} m_{\pi})^2
\nonumber \\
&=& 
- n! \left({- \langle \Sigma \rangle_{2 N} \over{2 m_{\pi}^{2}f_{\pi}^{2}}}
\right)^{n \over 2} (f_{\pi} m_{\pi})^2 
~.\
\label{e:sig4}
\end{eqnarray}
The iteration of such two-nucleon pion rescattering
diagrams, together with the vacuum term, leads once again to a 
geometric progression:
\begin{eqnarray} 
{\langle {\bar N} N \rangle_{\rho}
\over{\langle {\bar N} N \rangle_{0}}} \Bigg|_{\rm irr.}^{n-{\rm even}}
= {\langle {\bar q} q \rangle_{\rho}
\over{\langle {\bar q} q \rangle_{0}}} \Bigg|_{\rm irr.}^{n-{\rm even}}
&=&  
1 + \sum_{n = 2, 4, \cdots}^{\infty}   
\left({\rho \over{f_{\pi} m_{\pi}}}
\sqrt{- {1 \over 2} \langle \Sigma \rangle_{2 N}} 
\right)^{n}
\nonumber \\
&=&  
1 + \sum_{n = 1, 2, \cdots}^{\infty}   
\left({- \rho^{2} \langle \Sigma \rangle_{2 N} 
\over{2 (f_{\pi} m_{\pi})^{2}}} \right)^{n}
\nonumber \\
&=&  
\left[1 + {\rho^{2} \langle \Sigma \rangle_{2 N} 
\over{2(f_{\pi} m_{\pi})^{2}}} \right]^{-1}~. \
\label{e:ldt5}
\end{eqnarray}

(ii) For n-odd the amplitude is
in the form of a product of one $\pi N$, [or $\pi mN$ graph, with m-odd] 
and $(n - 1)/2$, [or $(n - m)/2$] two-nucleon irreducible graphs connected 
by $(n + 1)/2$ pion propagators.
Since one can select the single $\pi N$ graph to be any 
of the $n$ nucleon lines, and the first pair in 
$\left(\begin{array}{c} n - 1 \\ 2 \end{array} \right)$ many 
ways, the second pair in 
$\left(\begin{array}{c} n - 3 \\ 2 \end{array} \right)$ many 
ways, etc., one finds that
\begin{eqnarray}  
\langle n N \left| \Sigma \right| n N \rangle |^{n-{\rm odd}}
&=&   
n \left(\begin{array}{c} n - 1 \\ 2 \end{array} \right) 
\left(\begin{array}{c} n - 3 \\ 2 \end{array} \right) 
\cdots
\left(\begin{array}{c} 2 \\ 2 \end{array} \right) 
\langle \Sigma \rangle_{N} 
\left({- \langle \Sigma \rangle_{2 N} \over{m_{\pi}^{2} f_{\pi}^{2}}}
\right)^{n - 1 \over 2} 
\nonumber \\
&=& 
n! \langle \Sigma \rangle_{N} 
\left({- \langle \Sigma \rangle_{2 N} \over{2 f_{\pi}^{2} m_{\pi}^{2}}}
\right)^{n - 1 \over 2} ~.\
\label{e:sig5}
\end{eqnarray}
Hence we conclude
\begin{eqnarray} 
{\langle {\bar N} N \rangle_{\rho}
\over{\langle {\bar N} N \rangle_{0}}} \Bigg|^{n-{\rm odd}}
=
{\langle {\bar q} q \rangle_{\rho}
\over{\langle {\bar q} q \rangle_{0}}} \Bigg|^{n-{\rm odd}}
&=&  
- { \rho \langle \Sigma \rangle_{N} \over{(f_{\pi} m_{\pi})^{2}}} 
- \sum_{n = 3 , 5, \cdots}^{\infty}   
{ \rho \langle \Sigma \rangle_{N} \over{(f_{\pi} m_{\pi})^{2}}}
\left({\rho 
\over{f_{\pi} m_{\pi}}} 
\sqrt{- {1 \over 2} \langle \Sigma \rangle_{2 N}} 
\right)^{n - 1}
\nonumber \\
&=&  
-    
{\rho \langle \Sigma \rangle_{N} \over{(f_{\pi} m_{\pi})^{2}}}
\sum_{n = 0, 1, 2, ...}^{\infty}
\left({- \rho^{2} \langle \Sigma \rangle_{2 N} 
\over{2 (f_{\pi} m_{\pi})^{2}}} \right)^{n}
\nonumber \\
&=&  
- {\rho \langle \Sigma \rangle_{N} \over{(f_{\pi} m_{\pi})^{2}}}
\left[1 + 
{\rho^{2} \langle \Sigma \rangle_{2 N} \over{(f_{\pi} m_{\pi})^{2}}} 
\right]^{-1}
~. \
\label{e:ldt6}
\end{eqnarray}
Putting these two together we find
\begin{eqnarray} 
{\langle {\bar q} q \rangle_{\rho}
\over{\langle {\bar q} q \rangle_{0}}}
&=&  
{\langle {\bar q} q \rangle_{\rho}
\over{\langle {\bar q} q \rangle_{0}}} \Bigg|^{n-{\rm even}}
+ {\langle {\bar q} q \rangle_{\rho}
\over{\langle {\bar q} q \rangle_{0}}} \Bigg|^{n-{\rm odd}}
\nonumber \\
&=&  
\left[1 - {\rho \langle \Sigma \rangle_{N} \over{(f_{\pi} m_{\pi})^{2}}}\right]
\left[1 + 
{\rho^{2} \langle \Sigma \rangle_{2 N} \over{2 (f_{\pi} m_{\pi})^{2}}} 
\right]^{-1}~. \
\label{e:ldt7}
\end{eqnarray}
This result has two conjugate poles in the complex density plane at a distance 
$$\rho^2 \leq \rho_{c_{2}}^2 =  \left| {(f_{\pi} m_{\pi})^2 
\over{- \langle \Sigma \rangle_{2 N}}} \right|$$ 
from the origin. It is clear that the model- and approximation-dependent sign of 
$\langle \Sigma \rangle_{2 N}$ determines the position of the poles in the
complex $\rho$ plane, and in particular if they are on the real axis, or not. 
For instance in the Born approximation to the linear sigma model
$$ \langle \Sigma \rangle_{N} = g_{0} f_{\pi}
\left({m_{\pi} \over{m_{\sigma}}}\right)^{2}, $$
therefore
\begin{eqnarray}
\langle \Sigma \rangle_{2N} 
&=& 
- 2 \left({g_{0} m_{\pi} \over{m_{\sigma}^{2}}}\right)^{2}
+ 
3 \left({g_{0} m_{\pi} \over{m_{\sigma}^{2}}}\right)^{2}
\nonumber \\
&=& 
\left({g_{0} m_{\pi} \over{m_{\sigma}^{2}}}\right)^{2} 
\nonumber \\
&=& 
- {1 \over 2} \langle \Sigma \rangle_{2N}^{\pi}~. \
\label{e:sigma2nt2}
\end{eqnarray}
Thus in the linear sigma model $\langle \Sigma \rangle_{2 N}$ is positive,
as given in Eq. (\ref{e:sigma12}), and the poles are at imaginary densities
$\pm i \rho_{c_{2}}$ a factor $\sqrt{2}$ larger in absolute value
than $\rho_{c_{1}}$ predicted by the one-nucleon rescattering series. 

This sort of analysis can and ought to be extended to 3-, 4-, and higher
N-pion rescattering graphs. We close with a {\it conjecture}: In the linear 
sigma model the 3-nucleon critical density $\rho_{c_{3}}$ is larger than the 
two-nucleon one  $\rho_{c_{2}}$, the 4-N higher $\rho_{c_{4}}$ than the 3-N 
one, etc..

\section{Summary and conclusions}

In summary, in this paper we have: (i) given an explicit formula for the
n$^{th}$-order term in the powers-of-density ($\rho^n$) expansion of the 
nuclear matter 
sigma term $\langle \Sigma \rangle_{\rho}$ and of the nuclear matter 
fermion condensate $\langle {\bar q} q \rangle_{\rho}$ in particular; 
(ii) calculated the ${\cal O}(\rho^2)$ coefficient in the 
Born approximation to the linear sigma model; 
(iii) iterated these primitive ${\cal O}(\rho^2)$ terms to infinite order; 
(iv) found poles and radii of convergence associated with two such resummations. 
Thus we found that the ``larger'' of the two sums actually has the larger 
radius of convergence.

Perhaps the most important conceptual contribution of this paper 
is that of putting the older ``intuitive'' calculations onto formally 
sounder grounds of QFT. In particular, we have shown that 
the fundamental elements of a calculation of $\langle \Sigma \rangle_{\rho}$ 
are the connected n-nucleon sigma term matrix elements. 
All calculations of $\langle {\bar q} q \rangle_{\rho}$ can be reduced to these 
elements. Moreover, this result gives a clear definition
of the ``NN correlations'' in this context.

Some of the ideas introduced in this paper may have been tacitly assumed in 
earlier work, most notably that by the Manchester \cite{birse93} and the
Lyon groups \cite{eric93,lyon95}.
M. Ericson \cite{eric93} correctly concluded 
that the resummation of the single-nucleon sigma term leads to a geometric 
progression in the baryon density.
Similarly, Birse and McGovern \cite{birse93} came close to formulating the 
correct $O(\rho^2)$ prediction of the LDT in the linear sigma model.
Finally, some of the diagrammatic methods used here were developed  
in the 60's and 70's \cite{hub57,camb77,fw69}
for the description of pion propagation in nuclear matter. 

\acknowledgments

The author would like to thank K. Kubodera and F. Myhrer for valuable 
discussions and comments on the manuscript.

\widetext

\begin{figure}
\begin{center}
\epsfig{file=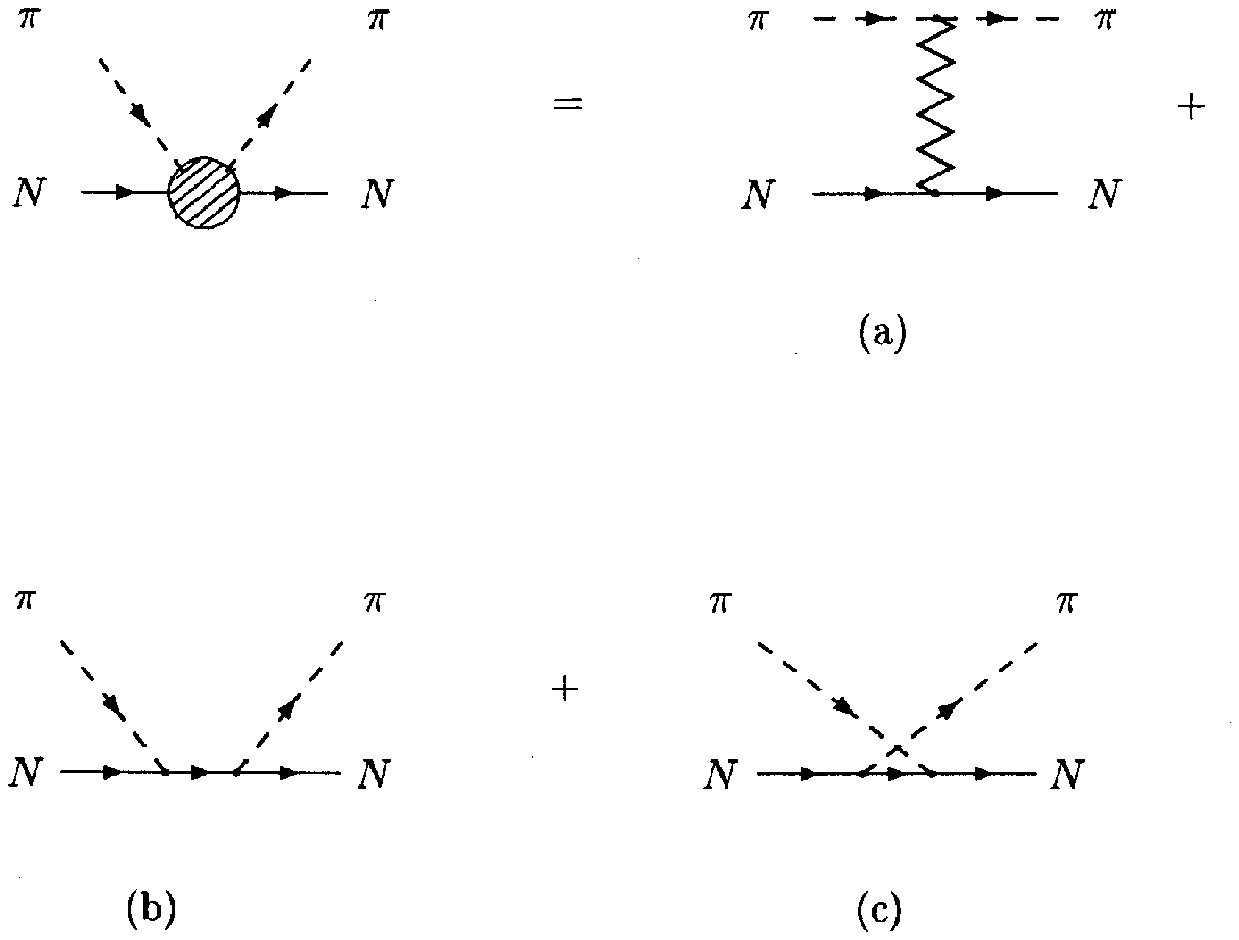,width=16cm} 
\end{center} 
\caption{ 
Feynman diagrams contributing to the ``elementary'' $\pi N$
elastic scattering amplitude: (a) the sigma-pole graph, and the (b), (c)
the nucleon-pole graphs. The zig-zag line denotes a sigma meson, the dashed
one a pion and the solid one a nucleon. 
\label{f:piN}}
\end{figure}

\begin{figure}

\begin{center}
\epsfig{file=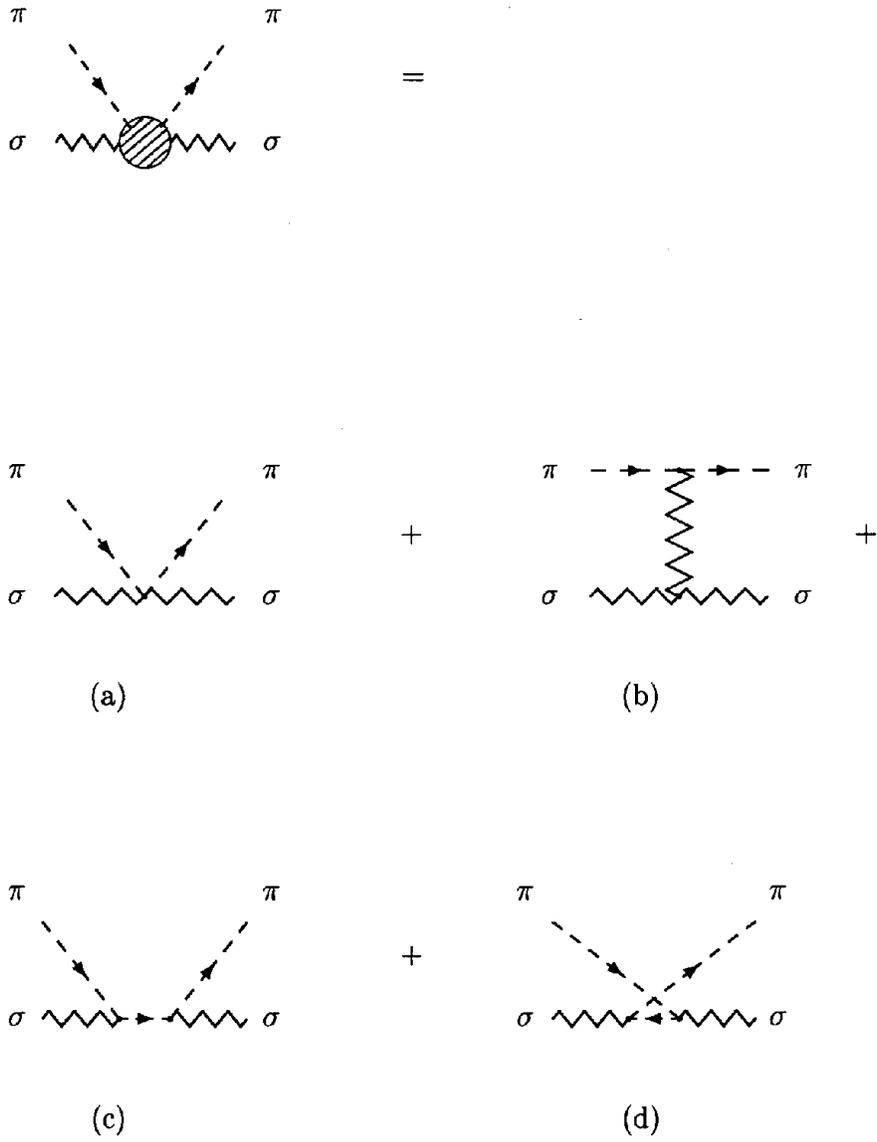,width=16cm}
\end{center} 
\caption{ 
Feynman diagrams contributing to the ``elementary'' $\pi \sigma$
elastic scattering amplitude: (a) and the (b), ie., to the sigma-meson's 
sigma term.
\label{f:pisig}}
\end{figure}

\begin{figure}
\begin{center}
\epsfig{file=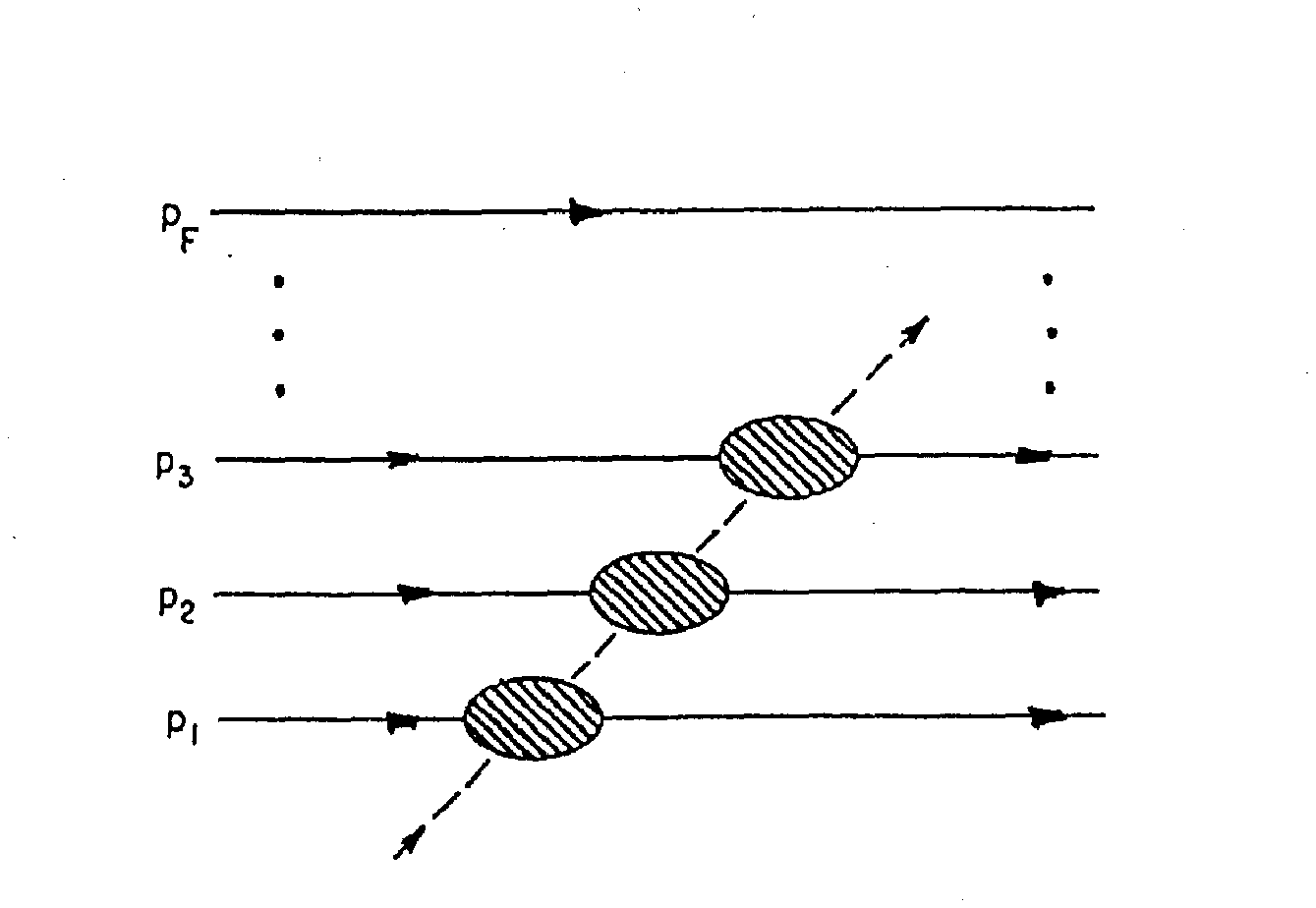,width=16cm} 
\end{center} 
\caption{ 
An example of order $n = 3$ reducible single-nucleon pion-rescattering graph.
\label{f:o3}}
\end{figure}

\begin{figure}
\begin{center} \epsfig{file=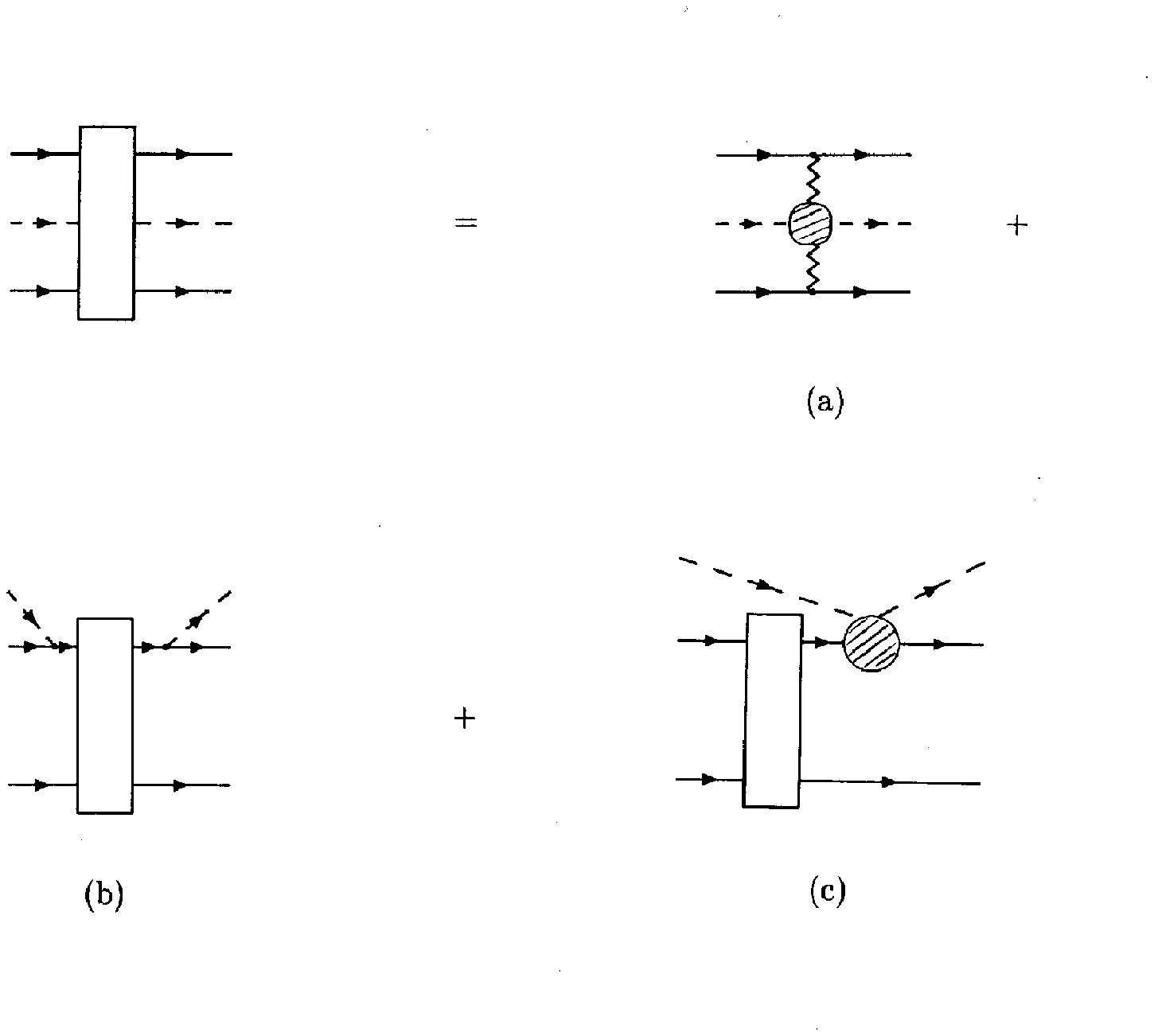,width=16cm} \end{center} 
\caption{ 
The pion-two-nucleon scattering amplitude: (a) the sigma-meson sigma term 
effective graph; (b) intermediate-state interaction graphs, (c) initial- and 
final-state interaction graphs. Graphs (b) and (c) may be termed 
$NN$ correlation effects. The square box with four external nucleon lines 
denotes and $NN$ potential due to the exchange of a single pion and sigma meson.
The hatched ``blob" in (a) represents the ``elementary'' $\pi N$
elastic scattering amplitude, such as the one shown in Fig. (1) in the linear 
sigma model. 
\label{f:2body}}
\end{figure}

\end{document}